\documentclass[prd,preprintnumbers,aps,preprint,amsmath,amssymb,superscriptaddress,floatfix,nofootinbib,unsortedaddress]{revtex4}
\pdfoutput=1

\usepackage{amssymb, bm, amsmath, graphicx, physymb}
\usepackage[colorlinks=true,linktocpage=true,linkcolor=blue,citecolor=blue]{hyperref}
\usepackage[usenames,dvipsnames]{color}
\usepackage{slashed}
\usepackage[utf8]{inputenc}
\usepackage[T1]{fontenc}
\usepackage{enumerate}
\usepackage{amsfonts}
\usepackage{mathtools}
\usepackage{latexsym}
\usepackage{hyperref}
\usepackage{epstopdf} % COMPILE EPS (AND PDF) FIGURES USING PDFLATEX!!!
\usepackage{bbm}
\usepackage{soul}
\usepackage[normalem]{ulem}
\usepackage{braket}
\usepackage{nccmath}
\usepackage{graphicx}
\usepackage{xcolor}

\usepackage[e]{esvect}

\def\cP{{\cal P}}
\def\cK{{\cal K}}

\def\cW{{\cal W}}

\newcommand{\secn}[1]{Section~1}
\newcommand{\appn}[1]{Appendix~1}

\long\def\comment#1{ }

\def\and{\quad\text{and}\quad}

\def\0{{\boldsymbol 0}}
\def\1{{\boldsymbol 1}}
\def\p{{\boldsymbol p}}

\def\x{{\boldsymbol x}}
\def\y{{\boldsymbol y}}

\def\Y{{\boldsymbol Y}}
\def\r{{\boldsymbol r}}

\def\w{{\boldsymbol w}}

\def\0{{\boldsymbol 0}}

\newcommand{\E}{\mcal{E}}
\renewcommand\a{\alpha}

\renewcommand\d{\delta}

\renewcommand\t{\tau}

\renewcommand\o{\omega}

\newcommand\m{\mu}

\renewcommand\E{{\bf E}}

\newcommand{\re}{{\rm{Re}}}

\newcommand{\tvec}{\boldsymbol}

\renewcommand{\part}{{\rm part}}
\newcommand{\be}{\begin{equation}}
\newcommand{\ee}{\end{equation}}
\newcommand{\bes}{\begin{subequations}}
\newcommand{\ees}{\end{subequations}}
\newcommand{\bea}{\begin{eqnarray}}
\newcommand{\eea}{\end{eqnarray}}

\newcommand{\pa}{\partial}

\newcommand{\na}{\nabla}
\newcommand{\beq}{\begin{equation}}
\newcommand{\eeq}{\end{equation}}

\makeatletter
\newsavebox{\@brx}
\newcommand{\llangle}[1][]{\savebox{\@brx}{\(\m@th{#1\langle}\)}%
  \mathopen{\copy\@brx\mkern2mu\kern-0.9\wd\@brx\usebox{\@brx}}}
\newcommand{\rrangle}[1][]{\savebox{\@brx}{\(\m@th{#1\rangle}\)}%
  \mathclose{\copy\@brx\mkern2mu\kern-0.9\wd\@brx\usebox{\@brx}}}
\makeatother

\begin{document}

\title{Jet quenching in the glasma phase: medium-induced radiation}

\begin{abstract}

Inspired by the recent considerations of parton momentum broadening in the glasma phase, we study the medium-induced soft gluon radiation of jet partons at early times in heavy-ion collisions. The glasma state is assumed to be comprised of independent color domains with homogenous longitudinal fields that vary event by event, and we further complete this model with an event-averaging procedure accounting for the finite correlation length. Using this description, we evaluate the rate of medium-induced radiation from an energetic parton at midrapidity in the glasma phase. We mainly focus on SU(2) color fields for simplicity, also referring to the U(1) case and comparing with the BDMPS-Z rate to gain further insight. Our results show that there is an intricate interplay of the synchrotron-like radiation in a single color domain with the destructive interference between different color domains, after the medium averaging is performed. Thus, we find that the emission rate is sensitive to the matter structure, decreasing for a glasma state populated by smaller color domains, i.e. for a glasma with a larger characteristic saturation scale. Our approach can be applied to more realistic backgrounds, and sets the stage for the modelling of jet evolution in the early stages of heavy-ion collisions.

\end{abstract}

\author{Jo\~{a}o Barata}
\email[Email: ]{jlourenco@bnl.gov}
\affiliation{Physics Department, Brookhaven National Laboratory, Upton, NY 11973, USA}

\author{Sigtryggur Hauksson}
\email[Email: ]{sigtryggur.hauksson@ipht.fr}
\affiliation{Institut de Physique Théorique, CEA/Saclay, Université Paris-Saclay, 91191, Gif sur Yvette, France}

\author{Xo{\'{a}}n Mayo L\'{o}pez}
\email[Email: ]{xoan.mayo.lopez@usc.es}
\affiliation{Instituto Galego de F{\'{i}}sica de Altas Enerx{\'{i}}as,  Universidade de Santiago de Compostela, Santiago de Compostela 15782, Galicia, Spain}

\author{Andrey V. Sadofyev}
\email[Email: ]{sadofyev@lip.pt}
\affiliation{LIP, Av. Prof. Gama Pinto, 2, P-1649-003 Lisboa, Portugal}

\maketitle

\section{Introduction}

Jets are essential probes of the nuclear medium formed in relativistic heavy-ion collisions and governed by quantum chromodynamics (QCD), for recent reviews see \cite{Cunqueiro:2021wls, Apolinario:2022vzg}. As a jet traverses the medium, it loses energy and has its substructure modified, carrying information about the matter properties along its path. Thus, jet observables can provide access to the details of the medium evolution, and jets may serve as a tomographic tool, see e.g. \cite{Vitev:2002pf, Wang:2002ri, JET:2013cls, Betz:2014cza, Xu:2014ica, Djordjevic:2016vfo, Apolinario:2017sob,He:2020iow,Sadofyev:2021ohn,Antiporda:2021hpk,Du:2021pqa} and references therein.

To accurately model jet modification in heavy-ion collisions, one needs a detailed understanding of the evolution of the QCD medium. This work focuses on the earliest stage just after the collision of the heavy ions, which has been argued to be characterized by strong (classical) color fields, initially aligned along the beam axis. This state of matter is usually referred to as the glasma. The glasma is formed when the color fields in each nucleus, sourced by partons in the color-glass condensate framework, start interacting just after the initial collision
\cite{Iancu_2004, Lappi_2006, Gelis_2010,  GELIS_2013, Gelis_2015}. In this regime, the matter has an intrinsic energy scale \(Q_s\), known as the saturation scale, which controls the energy density of the glasma. The saturation scale furthermore sets the correlation length of the color fields to be \(1/Q_s\). As a result, in the early stages of heavy-ion collisions, the QCD matter can be modeled as a collection of independent color domains with transverse size $1/Q_s$ within the glasma picture.

Due to high occupation density of the gluons, the glasma obeys classical equations of motion which are commonly solved on a discrete lattice, see e.g. \cite{Lappi_2003,Fukushima_2012,Berges_2014,Ipp_2018}. The glasma framework has been successfully used to describe final-state correlations of soft hadrons in heavy-ion collisions \cite{Schenke:2012wb, Schenke:2014tga, McDonald:2023qwc}. More recently, it has been extended to full (3+1)-D calculations, which do not assume boost invariance and have the potential to correctly predict correlations of final-state hadrons in rapidity \cite{Schenke:2016ksl, Gelfand_2016, Ipp_2020, Schlichting_2021, Ipp_2021, McDonald:2023qwc}. 

The glasma stage of heavy-ion collisions is short-lived, lasting less than $1$ fm/$c$. As the medium expands, the occupation density of gluons rapidly goes down and the medium can no longer be described classically. At this stage, the matter is still far from equilibrium, and its thermalization process can be described by the QCD kinetic theory, see e.g. \cite{Arnold_2003, Kurkela_2015, Kurkela_2019}. Once the system approaches the thermalization, the bulk of the evolution of the medium sets in: the quark-gluon plasma (QGP) described by the relativistic hydrodynamics has been formed. Until very recently\footnote{For a very early consideration of the jet quenching in glasma, see also \cite{Aurenche:2012qk}.}, nearly all considerations of jets in medium have  only included the QGP stage, and not the glasma and kinetic theory stages that precede it. The evolution of jets in the later stages of the medium evolution, when quarks and gluons have confined to hadrons, has also been given little theoretical attention, see e.g.~\cite{Dorau_2020}.

Even though the glasma phase is short-lived, the energy density of the medium is very high, meaning that the glasma has the potential to substantially modify jet substructure. One way to estimate the relative importance of the glasma stage is to measure its jet quenching parameter $\hat{q}$, the rate of the transverse momentum broadening of jet partons traversing it. This quantity has been measured in simulations of the glasma \cite{Ipp:2020mjc, Ipp:2020nfu, Avramescu:2023qvv}, as well as calculated analytically at very early times \cite{Carrington:2020sww,Carrington:2021dvw, Carrington:2022bnv}. Despite considerable theoretical uncertainty, these studies clearly indicate that \(\hat{q}\) in the glasma phase is very high. For instance, the results of \cite{Avramescu:2023qvv} suggest that \(\hat{q} \geq 5 \mathrm{GeV}^2/\mathrm{fm}\) during the first \(0.3\, \mathrm{fm}/c\) of the evolution, while the value extracted for the QGP phase at \(T = 200\,\mathrm{MeV}\) is \(\hat{q} \simeq 0.12\, \mathrm{GeV}^2/\mathrm{fm}\), see \cite{JETSCAPE:2021ehl}. In turn, the recent efforts to evaluate $\hat{q}$ during intermediate stages described by the kinetic theory also suggest rather large values of the jet quenching parameter \cite{Hauksson:2021okc,Boguslavski:2023alu, Boguslavski:2023waw,Boguslavski:2023jvg}, thus connecting the earlier- and later-time evolution. Interestingly, the momentum broadening in both the glasma and kinetic theory stages is anisotropic, potentially leading to more observable phenomena, such as the polarization of partons \cite{Kumar:2022ylt, Hauksson:2023tze} and photons \cite{Hauksson:2023dwh}.

However, it should be emphasized that more detailed studies are needed to quantify the importance of the early stages of the matter produced in heavy-ion collisions for in-medium evolution of jets. In particular, the high value of \(\hat{q}\) could be compensated by the short lifetime of the glasma phase, while its imprints may be washed out by the hydrodynamic evolution in the later stages. Nevertheless, it has been argued \cite{Carrington:2021dvw} that at the level of simple estimates, the contributions of the glasma and hydrodynamic stages to the total momentum broadening are similar.

Transverse momentum broadening of jet partons leads to medium-induced radiation, since on-shell jet partons can radiate gluons due to the interactions with the matter. The high rate of momentum broadening in the glasma suggests that such radiation should occur during the glasma stage too. However, only a few works have considered the medium-induced radiation at such early times, including \cite{Shuryak:2002ai,Zakharov:2008uk} which assumed a medium with a constant classical field and \cite{Aurenche:2012qk} that focused on a particular time-varying field without event-by-event averaging and spatial correlations. On the contrary, the medium-induced radiation has been studied in great detail in the QGP phase, and shown to modify the substructure of the jets compared with the vacuum case, leading to jet energy loss as soft gluons transport the jet energy down to the medium scale, see e.g. \cite{Mehtar-Tani:2022zwf,Blaizot:2013hx} and discussion therein. Moreover, there is an ongoing effort to improve the jet quenching theory in the QGP phase, e.g. by including the details of the medium evolution into the description \cite{He:2020iow,Sadofyev:2021ohn,Antiporda:2021hpk,Sadofyev:2022hhw,Barata:2022krd,Andres:2022ndd,Fu:2022idl,Barata:2022utc,Barata:2023qds,Kuzmin:2023hko} and by constructing tomographic observables sensitive to these effects \cite{Antiporda:2021hpk,Barata:2023zqg}. Thus, further improvements of the jet quenching theory in the neighboring phases are much needed, since only the overall effect of the matter on the jet observables is accessible experimentally.

Calculating the full spectrum of medium-induced emission in the glasma is a highly non-trivial task. An important feature of this stage of the matter evolution is the confinement of the gluon fields in domains of size $1/Q_s$ in the transverse plane.  Thus, the fields cannot be assumed to have only local correlations in time, i.e. the classical fields generated during this stage should not satisfy \(\langle A(t_1) A(t_2)\rangle \sim \delta(t_1 - t_2)\), where $t_1$ and $t_2$ are some positions along the jet path. Notice that this assumption drastically simplifies the jet quenching calculations in the QGP phase, where local correlations are typically assumed. A complete calculation of gluon radiation in the glasma thus demands the detailed event-by-event profile for the gluon fields and requires doing a quantum evolution of partons in that profile, see e.g. \cite{Li:2021zaw, Li:2023jeh, Barata:2023clv}. Another way to see the importance of the correlation length \(1/Q_s\), is to note that when the formation time for an emission is shorter than the correlation length in the glasma, the emission happens in a nearly constant field in contrast with the stochastic field picture underlying the jet quenching theory in the QGP phase. 

In this work, we simplify the problem of medium-induced gluon emission at early times by using a model that captures the essential features of the glasma. Our model assumes the medium to be comprised of color domains with sizes $\ell \sim 1/Q_s$, where each color domain has a constant chromoelectric field that varies event by event and is aligned along the beam axis. The fields in two color domains are independent of each other. This introduces a correlation length $\ell$ which relaxes the assumption of instantaneous/local correlations used in nearly all jet quenching calculations. Using this model, we obtain the rate of the momentum broadening and soft medium-induced radiation, both before and after averaging over events. For simplicity, we mainly consider a medium with SU(2) gauge fields, also relying on the U(1) case for illustrations, but our methods are generalizable to the physical SU(3) case. This approach could be directly applied to calculations with more realistic profiles: by dividing the full glasma profile in a single event up into slices with nearly constant fields, and averaging over different events, our model can approximate the full glasma.   

This paper is structured as follows. We start by discussing the single parton momentum broadening in Section~\ref{sec:broadening}, followed by a computation of the medium-induced soft gluon rate in Section~\ref{sec:radiation}. We first consider the case where the field fluctuations inside each color tube follow a Gaussian distribution, and verify our main conclusions within the same model completed with an alternative averaging procedure presented in Section~\ref{sec:delta_model}. In Section~\ref{sec:conclusion} we summarize our results and discuss possible future directions.

\section{Momentum broadening}\label{sec:broadening}

Focusing on jet quenching at early times of heavy-ion collisions, we use a simple physical picture for the glasma, taking into account the non-local correlations of color fields with correlation length \(\ell\sim 1/Q_s\) controlled by the saturation scale.  We refer to the regions where color fields are highly correlated as color domains, or alternatively as flux tubes, and view the medium as composed of multiple such flux tubes of a fixed size. In our model, the chromoelectric field is constant within the given flux tube, and aligned with the beam axis, as expected at very early times in heavy-ion collisions.

We define the coordinate system with respect to the jet axis so that $\x$ is transverse to the momentum of the leading parton, and assume that the jet is at midrapidity, so the beam axis is in the transverse direction, and the jet propagates along the $z$-axis. We write the field of our model as
\beq
\label{Eq:model}
A_{\mathrm{coh}}^{a\mu}(\x,z) = \d_\m^0 \, \x\cdot\E^a(z) = \left\{\begin{array}{lr} \delta^{\mu 0} \x\cdot \E^a_1, & 0\leq z < \ell  \\ \delta^{\mu 0} \x\cdot \E^a_2, & \ell \leq z < 2 \ell  \\ \delta^{\mu 0} \x\cdot \E^a_3, & 2\ell \leq z < 3\ell \\ \vdots  \end{array} \right.\,,
\eeq
where the first tube is assumed to start at $z=0$. This field is compatible with the Lorenz gauge with an additional axial condition $A_z=0$ which allows working with the physical polarizations of the gluon field. In what follows, without loss of generality, we will choose the beam direction as the $x$-axis, and, thus, keeping only the $x$-component of the electric field.\footnote{Notice, however, that the change of the field between two domains is assumed to be sufficiently slow comparing to the size of the domains, so that the $z$-component of the field can be neglected.}

As is usual in jet quenching considerations, all the results should be necessarily averaged over multiple events. We treat the electric fields in each flux tube as independent variables with a Gaussian distribution\footnote{Notice that a similar averaging procedure was used in \cite{Zakharov:2023oav} to describe the contribution of turbulent color fields to transverse momentum broadening.} of width $E_0$. Thus, for any function $f(E_{1x}^a,E_{2x}^a,E_{3x}^a,\cdots)$, the average over multiple events is
\begin{align}
\langle f(E_{1x}^a,E_{2x}^a,E_{3x}^a,\cdots) \rangle = \int_{E_1} e^{-E_{1x}^2/E_0^2} \int_{E_2} e^{-E_{2x}^2/E_0^2} \cdots f(E_{1x}^a,E_{2x}^a,E_{3x}^a,\cdots)\,,
\label{averaging1}
\end{align}
where $E_{nx}^2=\sum_a (E^a_{nx})^2$ in the case of SU($N_c$) fields, while $E_{nx}$ is just the $x$-component of the field in the given tube in the case of a U(1) background. Here, $n$ refers to the particular tube, $N_c$ is the number of colors, and we have introduced a shorthand notation: $\int_{E}=\int \frac{d^{N^2_c-1}E_x^a}{(\sqrt{\pi}E_0)^{N^2_c-1}}$ or  $\int_{E}=\int^\infty_{-\infty} \frac{dE_x}{(\sqrt{\pi}E_0)}$ for the SU($N_c$) and U(1) cases, respectively. Thus, our model only depends on two parameters, \(l\) and \(E_0\), after event averaging.

It is instructive to start with evaluating the transverse momentum broadening in our model. While the form of the medium-induced field differs from that in most BDMPS-Z considerations, see \cite{Casalderrey-Solana:2007knd} for a pedagogical review, the amplitudes  for transverse momentum broadening or gluon radiation can still be resummed. For instance, the jet quenching parameter for a parton of energy $\o$ in our model may be written as
\begin{align}
\hat{q}&=-\frac{1}{2(2\pi)^3\mathcal{N}}\frac{\pa}{\pa L}\int_{\x}\,\boldsymbol{\na}^2_{\x-\bar{\x}}\Big(J^\dag(\bar{\x})\mathcal{W}^\dag(\bar{\x})\mathcal{W}(\x) J(\x)\Big)_{\bar{\x}=\x}. 
\label{qhatWW}
\end{align}
where $\mathcal{N}=\frac{1}{2(2\pi)^3}\int_\p |J(\p)|^2$ is a normalization factor, and we keep track of how the sources $J(\x)=J^i(\o,\x)$ are distributed in the color space, as controlled by the superscript $i$. Notice that we have not yet averaged over multiple events, and this definition of $\hat{q}$ corresponds to the broadening in an ensemble of partons penetrating the same field configuration. 

For the color field in our model, the Wilson lines are given by $\cW(\x)\equiv\cW(\x;L,0)=\cP\exp\left\{i\int_0^L\,d\t\,t^a \E^a\cdot \x\right\}$ with $t^a$ being the generators of the particular representation of SU($N_c$), and assuming that the parton initial position $z_{in}$ coincides with the edge of the first tube, i.e. $z_{in}=0$. For convenience, we will assume that the energetic partons are in the fundamental representation. The Wilson lines are governed by a simple evolution equation
\begin{align}
&\frac{\pa}{\pa L}\cW(\x)=it^a \E^a(L)\cdot \x\,\cW(\x)\,,
\end{align}
which also controls their two-point functions
\begin{align}
&\frac{\pa}{\pa L}\cW(\x)\cW^\dag(\bar{\x})=i\E^a(L)\cdot\left[\x\,t^a\cW(\x)\cW^\dag(\bar{\x})-\bar{\x}\,\cW(\x)\cW^\dag(\bar{\x})t^a\right]\,.
\end{align}
Projecting the latter equation on the subspaces of the color structures, we find 
\begin{align}
&\frac{\pa}{\pa L}h^0=i\E^a(L)\cdot(\x-\bar{\x})h^a\notag\\
&\frac{\pa}{\pa L}h^a=i\E^b(L)\cdot\left[\frac{1}{2}(\x-\bar{\x})\left(\frac{\d^{ab}}{N_c}h^0+d^{abc} h^c\right)+\frac{i}{2}(\x+\bar{\x})f^{abc}h^c\right]\,,
\end{align}
where $h^0=\text{Tr}\,\cW(\x)\cW^\dag(\bar{\x})$ and $h^a=\text{Tr}\,t^a \cW(\x)\cW^\dag(\bar{\x})$. 

Since we are interested in $\hat{q}$ here, we have to compute only the particular path-length derivatives and transverse position gradients of $h$'s, and the problem can be further simplified. Indeed, 
\begin{align}
&\frac{\pa}{\pa L}{\boldsymbol\na}_{\y}\,h^0\big|_{\y=0}=0\,,\notag\\
&\frac{\pa}{\pa L}{\boldsymbol\na}_{\y}^2\,h^0\big|_{\y=0}=2i\E^a\cdot {\boldsymbol\na}_{\y}\,h^a\big|_{\y=0}\,,\notag\\
&\frac{\pa}{\pa L}{\boldsymbol\na}_{\y}\,h^a\big|_{\y=0}=i\left(\frac{\E^a}{2}+i\big(\E^b\cdot\Y\big)f^{abc}\,{\boldsymbol\na}_{\y}\,h^c\right)\big|_{\y=0}\,,\notag\\
&\frac{\pa}{\pa L}{\boldsymbol\na}_{\y}^2\,h^a\big|_{\y=0}=i\left(d^{abc}\,\E^b\cdot{\boldsymbol\na}_{\y}\, h^c+i\big(\E^b \cdot\Y\big)f^{abc}\,{\boldsymbol\na}_{\y}^2\,h^c\right)\big|_{\y=0}\,,
\end{align}
where $\Y=\frac{\x+\bar{\x}}{2}$ and $\y=\x-\bar{\x}$ are the relative coordinates. Here, we have also used that $h^a\big|_{\y=0}=0$ and $h^0\big|_{\y=0}=N_c$, since the Wilson lines cancel for $\y=0$. 
Furthermore, using that ${\boldsymbol\na}_{\y}^{n>0}\, h\big|_{L=0}=0$, since $\cW(\x)|_{L=0}=I$, one readily finds that
\begin{align}
&{\boldsymbol\na}_{\y}\,h^a\big|_{\y=0}=\frac{i}{2}\int^L_0 d\bar{\t}\,\tilde{\cW}^{ab}(\Y;L,\bar{\t})\E^b(\bar{\t})\,,
\end{align}
where $\tilde{\cW}^{ab}(\x;L,0)=\cP\exp\left(iT^c\int^L_0d\t\,\E^c(\t)\cdot\x\right)$ is the Wilson line in the adjoint representation, and $(T^c)_{ab}=-if^{abc}$ is the corresponding generator. The solutions for the second order transverse gradients of the color projections can be explicitly obtained, and read
\begin{align}
&{\boldsymbol\na}_{\y}^2\,h^0\big|_{\y=0}=- \int_0^{L} d\bar{\tau}\,\int^{\bar{\tau}}_0 d\t\;\tilde{\cW}^{ab}(\Y;\bar{\tau},\t) \; \E^a(\bar{\tau})\cdot\E^b(\t)\notag\\
&{\boldsymbol\na}_{\y}^2\,h^a\big|_{\y=0} = -\frac{1}{2} d^{ebc} \int^L_0 d\bar{\t}\int^{\bar{\t}}_0 d\t\,\tilde{\cW}^{ae}(\Y;L,\bar{\t}) \tilde{\cW}^{cd}(\Y;\bar{\t},\t)\;\E^b(\bar{\t})\cdot\E^d(\t)\,.
\label{ddh}
\end{align}
The first line measures total momentum broadening. The force exerted by the color field on the particle is integrated in time, both in the amplitude and in the conjugate amplitude.  The Wilson line inserted between the two fields ensures gauge invariance and is in the adjoint representation, like the gauge fields themselves. In the second line, we keep track of the final color of the particle. Thus, an additional Wilson line between the second field insertion and the measurement at time \(L\) is needed. 

One should notice here that $\hat{q}$ may in principle contain a contribution due to the octet component of the initial distribution. It may also have a non-trivial dependence on the spatial distribution of the initial ensemble of partons. However, to illustrate the physical picture, let us consider a simpler model for the initial distribution such that
\begin{align}
\frac{1}{2(2\pi)^3\mathcal{N}} \sum_{\text{sources}}J(\o,\p) J^\dag(\o,\bar{\p})=\frac{1}{2\pi w^2}\left(\mathbbm{1} + n^a t^a\right)e^{-\frac{\p^2+\bar{\p}^2}{4w^2}}
\end{align}
where we sum over possible initial sources, $n^a$ parameterizes the octet component\footnote{In the QGP phase the octet contribution to any jet observable is expected to vanish, see e.g.~\cite{Barata:2023uoi,Blaizot:2017ypk,Blaizot:1996az,Braaten:1990it}. However, in the presence of a coherent field, the octet contribution can survive, at least before averaging.}, and $w$ is the width of the initial distribution modeled by a Gaussian. Then, the jet quenching parameter can be written as 
\begin{align}
\hat{q}&=-\frac{w^2}{2\pi^3}\frac{\pa}{\pa L}\int_{\Y}\,{\boldsymbol\na}_{\y}^2\Big(h^0+n^ah^a\Big)_{\y=0} e^{-2w^2\Y^2}\,.
\end{align}
In what follows we will assume that the initial ensemble of jet partons has no net color for simplicity, setting $n^a=0$. Then, we can focus solely on the singlet part of the jet quenching process, and find
\begin{align}
\hat{q}(z)&=\frac{w^2}{2\pi^3}\int_{\Y}\int^z_0 d\bar{\t}\,\tilde{\cW}^{ab}(\Y;z,\bar{\t})\,E^a_x(z) E^b_x(\bar{\t})e^{-2w^2\Y^2}\,,
\label{qhatWEE}
\end{align}
where $z$ is the current position of the parton, and it is made explicit that only the $x$-component\footnote{In what follows, the $x$ subscript is dropped to alleviate the notation.} of the chromoelectric field is non-zero in the considered model. This form of $\hat{q}$ along with Eq.~\eqref{ddh} is in close relation with the earlier results for the transverse momentum broadening in a background field, {\it c.f.} \cite{Casalderrey-Solana:2007ahi,Majumder:2009cf%,Ipp:2020mjc
}.

The expression for $\hat{q}$ should be further averaged over multiple events, and we rely on the model in Eq.~\eqref{Eq:model}. While the piecewise form of the field is sufficiently simple, the averaging is still involved, and it is instructive to start with the simpler case of SU(2). Due to the path ordering, the Wilson line \(\tilde{\cW}(\Y;z,\bar{\tau})\) can be written as 
a product of Wilson lines for each flux tube
\begin{align}
\label{Eq:Wilson_product}
\tilde{\cW}(\mathbf{Y};z,\bar{\tau}) = \tilde{\cW}(\mathbf{Y};z,z_n)\,\tilde{\cW}(\mathbf{Y};z_n,z_{n-1})\, \tilde{\cW}(\mathbf{Y};z_{n-1},z_{n-2})  \cdots \tilde{\cW}(\mathbf{Y};z_{k+1},\bar{\tau}) 
\end{align}
where \(z_i = i l\) and \(z_{i-1}= (i-1) l\) are  the boundaries of the \(i\)th flux slab, and we have suppressed the color indices. In the first and the last Wilson lines, only a part of the flux tube is traversed, with \(z_{n+1} > z \geq z_n\) and \(z_{k+1} > \bar{\tau} \geq z_k\). The single tube Wilson lines are easily evaluated in SU(2), giving
\begin{align}
&\tilde{\cW}_{\text{SU(2)}}^{ab}(\Y;z_i,z_{i-1}) = \exp \left[ \epsilon^{abc} \, E^c_i Y \ell\right]
\notag\\&\hspace{1cm}
=\d^{ab}\cos{\big(E_iY\ell\big)}+2\frac{E^a_iE^b_i}{E^2_i}\sin^2\left(\frac{1}{2} E_iY\ell\right)+\epsilon^{abc}\frac{E^c_i}{E_i}\sin{\left(E_iY\ell\right)}\,,
\label{SU2Euler}
\end{align}
where \(\epsilon^{abc}\) is the structure constant of SU(2). Thus, unless both chromoelectric fields in Eq.~\eqref{qhatWEE} belong to the same flux tube, the average is zero either since the contributions are canceled by the antisymmetric structure constants or due to the absence of a preferred color-space direction in the averaging procedure. Consequently, we find that
\begin{align}
\langle\hat{q}(z)\rangle_{\text{SU(2)}}&= \frac{w^2}{2\pi^3}\int_{\Y}\int_E\int^z_{z_n} d\bar{\t} \,E^2\, e^{-2w^2\Y^2} e^{-\frac{E^2}{E_0^2}}=\frac{3E_0^2}{8\pi^2}\,(z-z_n)\,,
\end{align}
where $z_n$ is the last edge between two tubes penetrated by the parton when it is at $z$.

Turning to the physical case of SU(3), we immediately see that the same cancellation of the intermediate flux tubes is well expected. Indeed, the adjoint Wilson line $\tilde{W}(\Y;z,z_n)$ corresponding to the final tube can be expanded into a series, and all the terms except the zeroth-order one are transverse to $E^c(L)$. In turn, the zeroth-order term will average to zero due to the absence of any preferred direction in the color space, unless the other field is within the same tube. The resulting jet quenching parameter can be readily written as
\begin{align}
\langle\hat{q}(z)\rangle_{\text{SU(3)}}&= \frac{w^2}{2\pi^3}\int_{\Y}\int_E\int^z_{z_n} d\bar{\t}\, E^2 e^{-2w^2\Y^2} e^{-\frac{E^2}{E_0^2}}=\frac{E_0^2}{\pi^2}\,(z-z_n)\,.
\label{qhatscaling2}
\end{align}

Thus, we see that in our model, the jet quenching parameter is linearly increasing while the parton propagates along the given tube, and then falls to zero at the edge of the next tube, restarting the linear growth afterward. However, in any realistic setup, the initial position within the first flux tube, $z_{in}$, cannot be fixed, and one has to average over it. This is equivalent to averaging over the position within the last flux tube, $z-z_n$, and we readily find that
\begin{align}
\left\langle\hat{q}\right\rangle_{\text{SU(2)},\,z_{in}} = \frac{3E_0^2}{16\pi^2}\ell\,,~~~~~~~~~~~~\left\langle\hat{q}\right\rangle_{\text{SU(3)},\,z_{in}} = \frac{E_0^2}{2\pi^2}\ell\,,
\label{qhatscaling}
\end{align}
where the subscript $z_{in}$ indicates that the object has been additionally averaged over the initial position.

\section{Medium-induced radiation}\label{sec:radiation}

We now turn to medium-induced radiation in our color-domain model of the glasma. The underlying formalism is the same as in BDMPS-Z calculations, but since correlations of color fields are not localized in the glasma, the calculation is more involved qualitatively. In particular, the averaging of the final distribution is not fully factorized into the broadening and emission kernels as in BDMPS-Z, making it more challenging to evaluate. Therefore, we choose to focus on the radiation rate in the soft-gluon limit, which is insensitive to the broadening of the emitted gluon. This quantity can still be reduced to an emission kernel, and contains the relevant phenomenological information regarding the in-medium gluon production. 

The rate of the medium-induced radiation for an ensemble of partons propagating through the given field configuration is defined as
\begin{align}
   \frac{d\Gamma}{d\text{x}} =  \frac{2\alpha_s C_F}{\text{x}\o^2} \re \int_0^t ds\, \boldsymbol{\na}_\x\cdot \boldsymbol{\na}_\y \Big(\cK(\x,t;\y,s)- \cK_0(\x,t;\y,s)\Big)_{\x=\y=0}\,,
\end{align}
where \(\omega\) is the energy of the emitted gluon, x is its energy fraction with respect to the emitter, \(\cK\) is the in-medium emission kernel and $\cK_0$ is the emission kernel in vacuum. We have also assumed that the initial distribution is broad in momentum space $|J(\x)|^2\sim \d^{(2)}(\x-\x_0)$, and can be factorized. Notice that a change of the center of the initial distribution \(\x_0\) corresponds to a shift of the color field in Eq.~\eqref{Eq:model} by \(\x_0 \cdot \mathbf{E}_i^a\) in each flux tube, which amounts to a gauge transformation. Gauge invariance, therefore, allows us to set  $\x_0=0$ without loss of generality. Thus, the in-medium emission kernel is given by
\begin{align}
\cK(\x,t;\y,s)=\frac{1}{N_c^2-1} G^{ab}(\x,t;\y,s)\tilde{\cW}^{\dag ba}(0;t,s)
\end{align}
where \(G^{ab}\) is the single-particle (BDMPS-Z) propagator of the emitted gluon. More explicitly, the kernel can be expressed as a path integral, where the Wilson line accounts for the precession in the background color field
\begin{align}
&\cK(\x,t;\y,s)=\frac{1}{N_c^2-1}\int_{\mathbf{r}(s) = \mathbf{y}}^{\mathbf{r}(t) = \mathbf{x}} \mathcal{D}\mathbf{r}\,  \exp \left[ i \frac{\omega}{2} \int_s^{t} d\t\;   \dot{\tvec{r}}^2 \right]\notag\\
&\hspace{6cm}\times\text{Tr}\,\cP\exp \left[iT^c\int_s^t d\t\, \E^c(\t)\cdot \tvec{r}(\t)\right]\,.
\end{align}

\subsection{Induced radiation in QED}

The averaging procedure of our model is more involved, making it instructive to start with a toy consideration. For that purpose, let us focus on the U(1) case, which can be treated analytically. Then, the color structure becomes trivial
\begin{align}
\cK_{\text{U(1)}}(\x,t;\y,s)=\int \mathcal{D}\mathbf{r}\,  \exp \left[ i\int_s^{t} dt\;   \left(\frac{\omega}{2}\dot{\tvec{r}}^2+\E(\t)\cdot \tvec{r}(\t)\right)\right]\,,
\end{align}
and the path integral can be straightforwardly evaluated, see e.g. \cite{Aurenche:2012qk},
\begin{align}
&\cK_{\text{U(1)}}(\x,t;\y,s)=\frac{\omega}{2\pi i (t-s)}\exp\Bigg\{i \frac{\omega}{2} \frac{(\x-\y)^2}{t-s}\notag\\
&\hspace{1cm}+ \frac{i}{(t-s)}\left(\x\cdot\int_s^t d\t (\t-s)\E(\t)+\y\cdot\int_s^t d\t (t-\t)\E(\t)\right)\notag\\
&\hspace{2cm}  -  \frac{i}{\omega(t-s)}\int_s^t d\t\int_s^\t d\bar{\t}\,(t-\t)(\bar{\t}-s)E(\bar{\t})E(\t)\Bigg\}\,.
\label{Eq:U1pathintegral}
\end{align}
This form of the kernel is valid for general background field \(\mathbf{E}(\tau)\), and we further focus on the piecewise case of our matter model. 

The special case where emissions take place inside a single flux tube is particularly illustrative. For a single tube, the field \(\mathbf{E}\) is constant, and the kernel is translationally invariant in the longitudinal direction, simplifying to
\begin{align}
     &\cK_{\text{U(1)}}(\x,t;\y,0)=\frac{\omega}{2\pi i t} \exp\Bigg\{i \Bigg( \frac{\omega}{2t} (\x-\y)^2 + \frac{\E t}{2}\cdot(\x+\y)  -  \frac{E^2}{24 \omega}t^3\Bigg)\Bigg\}\,.
     \label{KU1onetube}
\end{align}
Thus, one may readily write the medium-induced radiation rate for constant background field \cite{Baier:1968,Zakharov:2008uk,Shuryak:2002ai, Zakharov:2016mmc} as
\begin{align}
    \frac{d\Gamma_{\text{U(1)}}}{d\text{x}} &= \frac{2\alpha_s C_F}{\text{x}\pi} \re \int_0^t ds\,  \frac{1}{s^2} \left(1 - \left(1- i\frac{E^2s^3}{8\omega} \right) e^{-i \frac{E^2 s^3}{24 \omega}}  \right)\,.
    \label{K1tubeNA}
\end{align}
This regime, corresponding to the field being constant during the emission process, is realized when the correlation length \(l \sim 1/Q_s\) is longer than the formation time.

Before averaging over the field configuration using Eq.~\eqref{averaging1}, let us consider the rate in the given event for the case of a single flux tube. At earlier times, the rate in Eq.~\eqref{K1tubeNA} grows as a power of the traveled distance: $\frac{d\Gamma_{\text{U(1)}}}{d\text{x}} \simeq \frac{7\alpha_s C_F}{24 \text{x}\pi}\frac{E^4}{\omega^2}\frac{t^5}{5!}$. At later times, it saturates, $\frac{d\Gamma_{\text{U(1)}}}{d\text{x}}\big|_{t\to \infty} = 3^{1/6}\, \Gamma \left(\frac{2}{3}\right)\,\frac{\alpha_s C_F}{\text{x}\pi}\, E^{2/3}\omega^{-1/3}$, after a characteristic time $t_{\rm ch}=\left(24\omega/E^2\right)^{1/3}$, which can be understood as the in-medium formation time. Indeed, the same parametric form follows from the scaling \(t_{\mathrm{ch}} \sim \sqrt{\omega/\hat{q}(t_{\mathrm{ch}})}\), appearing in the BDMPS-Z considerations \cite{Casalderrey-Solana:2007knd}, if the form of the jet quenching parameter in Eq.~\eqref{qhatscaling2} is used for a single long tube, i.e. \(\hat{q}(t_{\mathrm{ch}}) \sim E^2 t_{\mathrm{ch}}\). Thus, the in-medium formation time set by \(t_{\mathrm{ch}}\) for the single-tube case is (parametrically) shorter than the correlation length \(1/Q_s\) only for very soft gluons, such that \(\omega < Q_s/24\) when \(E \sim Q_s^2\).

\begin{figure}[t!]
    \centering
    \includegraphics[width=1\textwidth]{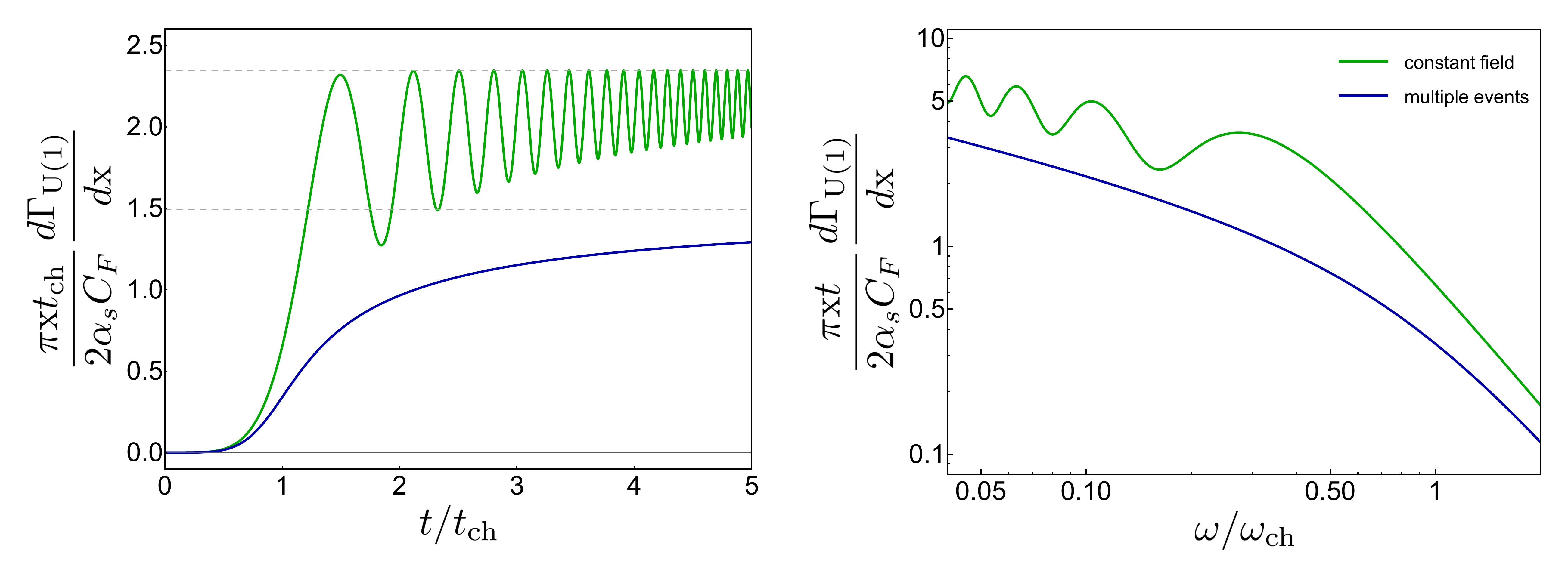}
    \caption{The rate of the medium-induced radiation for the background U(1) field is plotted for the constant field case (green) and Gaussian averaging over multiple events (blue), assuming a single long flux tube. Dashed lines indicate the corresponding asymptotic values. The characteristic time $t_{\text{ch}}$ and the characteristic frequency $\o_{\text{ch}}$ are defined with respect to $E$ (green) and $E_0$ (blue). 
    }
    \label{fig:U1rate1tube}
\end{figure}

One may further notice that the rate in Eq.~\eqref{K1tubeNA} takes a particularly simple form if we scale it with the 
characteristic time, and its parametric dependence is captured by a single function, $t_{\text{ch}}\frac{d\Gamma_{\text{U(1)}}}{d\text{x}} = f\left(t/t_{\text{ch}}\right)$. Alternatively, the parametric dependence of the rate can be re-expressed as a function of gluon energy $t \frac{d\Gamma_{\text{U(1)}}}{d\text{x}}=\tilde{f}(\omega/\omega_{\text{ch}})$ with $\omega_{\mathrm{ch}} = E^2t^3/24$. The two functions are directly related since $\omega/\omega_{\mathrm{ch}} = (t/t_{\mathrm{ch}})^{-3}$, and we show their behavior in Fig.~\ref{fig:U1rate1tube}. Notice that energetic gluons $\omega > \omega_{\mathrm{ch}}$ have barely had time to form by the measurement moment $t<t_{\mathrm{ch}} $, and the spectrum rapidly decays with $\o/\o_{\mathrm{ch}}$.

We now turn to the event-averaged results, still in the case of a single long tube. Averaging the emission kernel, Eq.~\eqref{KU1onetube}, over multiple events, we get 
\begin{align}
      &\left\langle\boldsymbol{\na}_\x\cdot \boldsymbol{\na}_\y\Big(\cK_{\text{U(1)}}(\x,t;\y,0)-\cK_{0}(\x,t;\y,0)\Big)_{\x=\y=0}\right\rangle\notag\\
      &\hspace{2cm}= \frac{\omega^2}{\pi t^2}\left[\frac{1}{\sqrt{i\frac{E_0^2t^3}{24 \omega}+1}}\left(-1+\frac{3}{2\left(1-i\frac{24\o}{E_0^2t^3}\right)}\right)+1\right] \,,
\end{align}
where the Gaussian integration in Eq.~\eqref{averaging1} has been performed explicitly. Then, the averaged rate of the medium-induced radiation is given by
\begin{align}
   &\left\langle\frac{d\Gamma_{\text{U(1)}}}{d\text{x}}\right\rangle =  \frac{2\alpha_s C_F}{\pi \text{x}} \re \int_0^t \frac{ds}{s^2}\,\left[ \frac{1}{\sqrt{i\frac{E_0^2s^3}{24 \omega}+1}}\left(-1+\frac{3}{2\left(1-i\frac{24\o}{E_0^2 s^3}\right)}\right)+1\right]\,.
   \label{U1singletube}
\end{align}
Taking the same limits as before, one can see that at earlier times the averaged rate scales as $\left\langle\frac{d\Gamma_{\text{U(1)}}}{d\text{x}}\right\rangle \simeq \frac{7\alpha_s C_F}{32\text{x}\pi}\frac{E_0^4}{\o^2}\frac{t^5}{5!}$, while at later times it tends to $\left\langle\frac{d\Gamma_{\text{U(1)}}}{d\text{x}}\right\rangle\big|_{t\to \infty} = 3^{1/6}\, \Gamma \left(\frac{2}{3}\right)\Gamma \left(\frac{5}{6}\right)\,\frac{\alpha_s C_F}{\text{x}\pi^{3/2}}\,   E_0^{2/3}\omega^{-1/3}$. All these characteristic features of the single flux tube case are summarized in Fig.~\ref{fig:U1rate1tube}.

The structure of the rate is more involved for the more realistic case when the leading parton traverses multiple tubes. Redefining the characteristic time $t_{\mathrm{ch}} = \left(24\omega/E_0^2\right)^{1/3}$ with respect to $E_0$ of the averaging procedure in Eq.~\eqref{averaging1}, one sees that the full multi-tube rate can be written simply as $t_{\rm ch}\frac{d\Gamma_{\text{U(1)}}}{dx}=t_{\rm ch}\frac{d\Gamma_{\text{U(1)}}}{dx}\left(t/t_{\mathrm{ch}}, \ell/t, \E_i/E_0\right)$. Averaging over the fields in each flux tube, the rate further simplifies, and its parametric dependence is captured by a single function $t_{\rm ch}\left\langle\frac{d\Gamma_{\text{U(1)}}}{dx}\right\rangle = F(t/t_{\mathrm{ch}}, \ell/t)$, which depends solely on the ratios $t/t_{\mathrm{ch}}$  and \(\ell/t\). Finding analytic results for $F$ quickly becomes unwieldy when \(\ell/t\) is small, i.e. when the radiation occurs over multiple flux tubes, requiring  multiple Gaussian integrals in the averaging procedure. However, with the explicit form of Eq.~\eqref{Eq:U1pathintegral} in hand, one can readily evaluate any finite number of Gaussian integrations over the field in Eq.~\eqref{averaging1} numerically.

\begin{figure}[t]
    \centering
    \includegraphics[width=\textwidth]{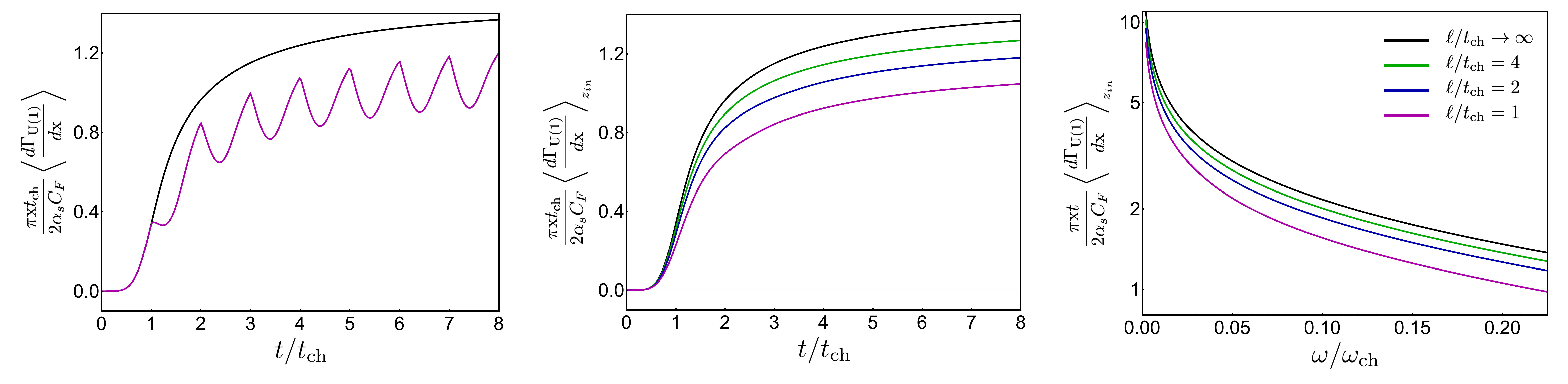}
    \caption{An illustration of the U(1) averaged rate. The left panel gives $\frac{\pi \text{x}}{2\a_s C_f}t_{\rm ch}\left\langle\frac{d\Gamma_{\rm U(1)}}{dx}\right\rangle$ with the leading parton starting from the edge of the first tube $z_{in}=0$. The middle and right panels depict $\frac{\pi \text{x}}{2\a_s C_f}t_{\rm ch}\left\langle\frac{d\Gamma_{\rm U(1)}}{dx}\right\rangle_{z_{in}}$ and $\frac{\pi x}{2\a_s C_f}t\left\langle\frac{d\Gamma_{\rm U(1)}}{dx}\right\rangle_{z_{in}}$ averaged over the starting position within the first tube. The colors correspond to the rate evaluated for different tube sizes, including the long tube limit (black), which appears in all three panels and is given by Eq.~\eqref{U1singletube}.}
    \label{fig:U1rate}
\end{figure}

To illustrate the features of the spectrum, we present numerical results for $\left\langle\frac{d\Gamma_{\text{U(1)}}}{dx}\right\rangle$ and $\left\langle\frac{d\Gamma_{\text{U(1)}}}{dx}\right\rangle_{z_{in}}$ in Fig.~\ref{fig:U1rate}. We find that, while the leading parton is traversing the first tube, the rate follows the single-tube case. Upon crossing the border between two tubes, the rate experiences some destructive interference, falling after that point. However, propagating further within each given tube, the rate builds up resembling its behavior in the first tube. This series of oscillations of the rate continues, staying under the single-tube curve. The peaks of the rate occur at the edges of the tubes, as depicted in the left panel of Fig.~\ref{fig:U1rate}. If the starting position $z_{in}$ within the first tube is varied, the rate curve gets modified, yet the overall pattern remains the same with the minima and peaks shifted. Averaging over the initial position of the leading parton in the first flux tube, we see that the series of minima and peaks is smeared, and the rate takes a shape similar to the constant field case. The greater the number of tubes traversed by the leading parton by the given $t/t_{\rm ch}$ the lower the averaged rate at that given point, and the curves for different $\ell/t_{\rm ch}$ appear to be ordered, as depicted in the middle panel of Fig.~\ref{fig:U1rate}. Finally, we find it instructive to plot the rate $t\left\langle\frac{d\Gamma_{\text{U(1)}}}{dx}\right\rangle_{z_{in}}$ as a function of frequency $\o/\o_{\rm ch}$,  as shown in the right panel of Fig.~\ref{fig:U1rate}, and the curves are ordered in the same way. Notice that the averaged rate as a function of $\o/\o_{\rm ch}$ is independent of the particular position of measurement $t$ when scaled with it.

The ordering of the rate curves can also be understood at the level of the broadening pattern. Indeed, in the case of shorter flux tubes, accumulating transverse momentum is harder, as the fields in different tubes can oppose each other. This results in a longer in-medium formation time\footnote{Notice that $t_{\mathrm{ch}}$ is independent of the number of flux tubes, while the physical in-medium formation time does depend on $\ell$. In the case of constant field, corresponding to large $\ell$, the two times coincide, but for shorter flux tubes the in-medium formation time is longer than the characteristic time.} for the emission process, leading to a lower rate. This effect is stronger for shorter tubes, reducing the rate. It is furthermore particularly important at larger measurement times \(t/t_{\mathrm{ch}}\), when the system has had sufficient time to propagate through multiple flux tubes, and at higher energies, where the formation time is longer and, thus, more flux tubes are traversed during the radiation process. 

\begin{figure}
    \centering
    \includegraphics[width=1\textwidth]{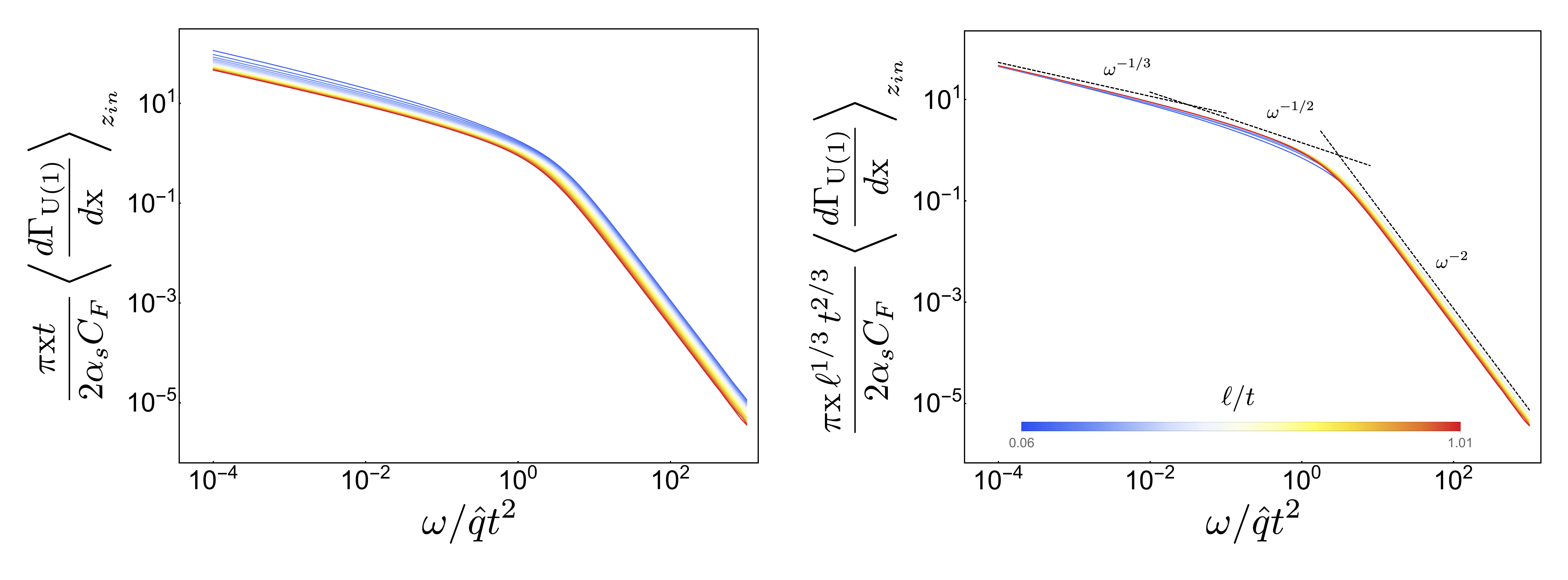}
    \caption{
    An illustration of the U(1) rate at constant $\hat{q}t^2$.
    The left panel depicts $\frac{\pi \text{x}\,t}{2\a_s C_f}\,\left\langle\frac{d\Gamma_{\rm U(1)}}{dx}\right\rangle_{z_{in}}$ as a function of $\o/\hat{q}t^2$, the right panel depicts its rescaled value $\frac{\pi \text{x}\,\ell^{1/3}\,t^{2/3}}{2\a_s C_f}\,\left\langle\frac{d\Gamma_{\rm U(1)}}{dx}\right\rangle_{z_{in}}$. The color of the solid lines corresponds to different values of $\ell/t$ as indicated in the legend, while the black dashed lines denote the asymptotics, emphasizing the local scaling of the rate with powers of $\o$.
    }
    \label{fig:fixedqhat}
\end{figure}

Finally, one may find it instructive to consider the parametric dependence of the rate in the regime where $\hat{q}$ is kept constant. Indeed, the BDMPS-Z formalism is often considered in the so-called harmonic approximation, see e.g. \cite{Casalderrey-Solana:2007knd}, which is fully controlled by $\hat{q}$, and one may wonder if it is the case in our model. Furthermore, the late-time behavior of the rate at fixed jet quenching parameter, $\langle\hat{q}\rangle_{\rm U(1)}=\frac{1}{8\pi^2}E_0^2\ell$, and varying correlation length provides an opportunity to study the dependence of the in-medium formation time on $\ell$. In the regime of interest, the parametric dependence of the rate can be summarized as follows: $t\left\langle\frac{d\Gamma_{\text{U(1)}}}{dx}\right\rangle_{z_{in}}\sim(t/\ell)^{1/3}(\hat{q}t^2/\o)^{1/3}F(\o/(\hat{q}t^2),\ell/t)$, and we illustrate it in the left panel of Fig.~\ref{fig:fixedqhat}. Strikingly, we find that the dependence of the rate on $\ell/t$ is dominated by the prefactor $(t/\ell)^{1/3}$ as can be seen from the right panel of Fig.~\ref{fig:fixedqhat}, suggesting that the in-medium formation time scales as $t_m/t \sim (\ell/t)^{1/3}$ at constant $\o/(\hat{q}t^2)$. Notice that the weaker dependence of the function $F$ on the correlation length is merely an observation, and further understanding of that behavior is required.

Turning to the energy dependence of the rate in Fig.~\ref{fig:fixedqhat}, we readily notice that there are three distinct scaling regimes. At low energies the formation time is small, the partons only traverse a few flux tubes while radiating, and the prefactor $(\hat{q}t^2/\o)^{1/3}$ dominates. This regime has the same energy scaling as in the case of a single flux tube, and the formation time parametrically follows the characteristic time $t_m/t \sim (\ell/t)^{1/3}(\o/(\hat{q}t^2))^{1/3}\sim t_{\rm ch}/t$. In the transition region of Fig.~\ref{fig:fixedqhat}, at larger energies, the scaling of the rate can be well approximated with $(\hat{q}/\o)^{1/2}$, resembling the well-known behavior of the BDMPS-Z rate under the harmonic approximation, and the formation time scales as $t_m/t \sim (\ell/t)^{1/3}(\o/(\hat{q}t^2))^{1/2}$. This is well expected since for sufficiently high gluon energies the formation time is larger than $\ell$, and the partons traverse multiple flux tubes, receiving stochastic transverse momentum kicks with nearly local correlations. This picture closely follows the assumption used in the BDMPS-Z formalism, recovering it in this parametric region. At even higher energies, the formation time for the emission is larger than the measurement time \(t\), the partons barely have time to radiate, and, consequently, the rate steeply falls, scaling as $1/\o^{2}$. This scaling, also appearing in the BDMPS-Z considerations, can be attributed to the limitations of the harmonic approximation, which cannot capture the Coulomb-type interactions, resulting in a less steep decline, see e.g. \cite{Barata:2020sav,Isaksen:2022pkj}. However, our model includes no such hard transverse momentum exchanges in its current form, and, in general, their presence and importance in the glasma phase should be further investigated. We leave this question of such potential corrections to the medium model at larger energies for future work.

\subsection{Induced radiation for SU($2$)}

Although the general case of SU($N_c$) is considerably more involved, some intuition regarding the non-trivial color dynamics can be gained by considering SU(2) background, thanks to the relatively simple form of its general group element. This is particularly relevant as many glasma simulations are based on SU(2) field theory, for a review see e.g. \cite{Berges:2020fwq}.

In our model, characterized by a piecewise field, the path ordered exponential can be expressed as an ordered product of simple matrix exponentials, see Eq.~\eqref{Eq:Wilson_product}. Each of these can then be expanded within the same matrix basis using Eq.~\eqref{SU2Euler}. Averaging the rate over the orientation of the fields in the color space, one may notice that each such matrix exponential can be treated independently. Indeed,
\begin{align}
\left\langle\tilde{\cW}^{ab}_{\text{SU(2)}}(\r; z_i, z_{i-1})\right\rangle = \left\langle\,\frac{1}{3}\d^{ab}\left\{2\cos{\left(E_i\int^{z_i}_{z_{i-1}} r_x(\t)d\t\right)}+1\right\}\right\rangle
\label{WSU2averaged}
\end{align}
while for the first and last flux tubes, the limits should be adjusted accordingly. Consequently, all the color structures in the kernel become trivial, and we can express the SU(2) kernel as a convolution of U(1) kernels before averaging the full expression over \(E_i\).

To illustrate that, let us first focus on the case of a single flux tube. Then, we can express the full SU(2) kernel as
\begin{align}
\left\langle\cK_{\text{SU(2)}}(\x,t;\y,0)\right\rangle=\frac{1}{3}\left\langle\,\cK_+(\x,t;\y,0)+\cK_-(\x,t;\y,0)+\cK_0(\x,t;\y,0)\,\right\rangle_{{\rm SU}(2)}\,,
\label{KSU2inKU1}
\end{align}
where $\cK_+$, $\cK_-$, and $\cK_0$ are the U(1) kernels for a single flux tube with electric field $E$, $-E$, and $E=0$ respectively, with $E>0$ since the absolute value has been taken in the color space. However, one should notice that the averaging in Eq.~\eqref{KSU2inKU1} is for the SU(2) field strength $E$, and the residual radial integral $\int_{0}^{\infty}\frac{4\pi E^2 dE}{\pi^{3/2}E_0^3}$ incorporates a Jacobian from the full color averaging. Therefore, we keep the corresponding subscript on the averaging to indicate that point.

Noticing that the expression in Eq.~\eqref{KSU2inKU1} is even under $E\to-E$, one may further write
\begin{align}
\left\langle\,\cK_+(\x,t;\y,0)+\cK_-(\x,t;\y,0)\right\rangle_{SU(2)}=\frac{4}{E_0^2}\left\langle\,E^2\,\cK_+(\x,t;\y,0)\right\rangle_{{\rm U}(1)}\,,
\label{KSU2inKU1_simpl}
\end{align}
where the $E$-integration in the r.h.s. runs from $-\infty$ to $\infty$. In what follows, we will omit the U(1) subscript in the average of $\cK_\pm$. Then, the averaged SU(2) radiation rate for the case of a single flux tube can be explicitly written as
\begin{align}
   &\left\langle\frac{d\Gamma_{\text{SU(2)}}}{d\text{x}}\right\rangle =  \frac{4\alpha_s C_F}{3\text{x}\o^2} \re \int_0^t ds\,\boldsymbol{\na}_\x\cdot \boldsymbol{\na}_\y \left(\frac{2}{E_0^2}\left\langle\,E^2\cK_+(\x,t;\y,s)\right\rangle-\cK_0(\x,t;\y,s)\right)_{\x=\y=0}\notag\\
   &\hspace{1.5cm}=\frac{4\alpha_s C_F}{3\pi\text{x}} \re \int_0^t \frac{ds}{s^2} \left( \frac{1}{\left(i\frac{E_0^2s^3}{24\omega}+1\right)^{\frac{3}{2}}}\left(-1+\frac{9}{2\left(1-i\frac{24\o}{E_0^2s^3}\right)}\right)+1\right)\,.
   \label{SU2singletube}
\end{align}
Similarly to the U(1) case, in the early time limit $\frac{d\Gamma_{\text{SU(2)}}}{d\text{x}} \sim t^5$, while at later times it scales with the inverse of the characteristic  time $\left\langle\frac{d\Gamma_{\text{SU(2)}}}{d\text{x}}\right\rangle\big|_{t\to \infty} \sim 1/t_{\rm ch}$.

With this simplification of the color structure of the averaged kernel, the case of multiple flux tubes can be treated in the same way as in the U(1) case. For instance, for $t<2\ell$ we can readily write
\begin{align}
   \left\langle\frac{d\Gamma_{\text{SU(2)}}}{d\text{x}}\right\rangle &=  
   \frac{2\alpha_s C_F}{\text{x}\o^2} \re \int_0^t ds\, \boldsymbol{\na}_\x\cdot \boldsymbol{\na}_\y \Big(\left(\theta(\ell-t)+\theta(s-\ell)\right)\left\langle\cK^{(1)}_{\text{SU(2)}}(\x,t;\y,s)\right\rangle\notag\\
   &\hspace{-0.5cm}+\theta(t-\ell)\theta(\ell-s)\int_\w\left\langle\cK^{(1)}_{\text{SU(2)}}(\x,t;\w,\ell)\right\rangle\left\langle\cK^{(1)}_{\text{SU(2)}}(\w,\ell;\y,s)\right\rangle-\cK_0(\x,t;\y,s)\Big)_{\x=\y=0}\,,
\end{align}
where all the averagings run over SU(2) fields, and the superscript indicates that the particular kernels are taken for the single tube (or, equivalently, constant field) case as given by Eq.~\eqref{KSU2inKU1}. One may also notice that the parameter dependence of the SU(2) rate can be treated in the same way as in the U(1) case, rescaling it with $t_{\rm ch}$ or $t$, and focusing on its dependence on $t/t_{\rm ch}$ or $\o/\o_{\rm ch}$ with the characteristic scales being the same.

Thus, obtaining the SU(2) kernel for a few tubes in a closed analytical form is sufficiently straightforward. However, the Gaussian integrals of the averaging procedure still need to be treated numerically. Additionally, the need for the integrations over the intermediate transverse positions at the edges of the tubes complicates the numerics, and we focus on a smaller number of tubes.

\begin{figure}[t]
    \centering
    \includegraphics[width=\textwidth]{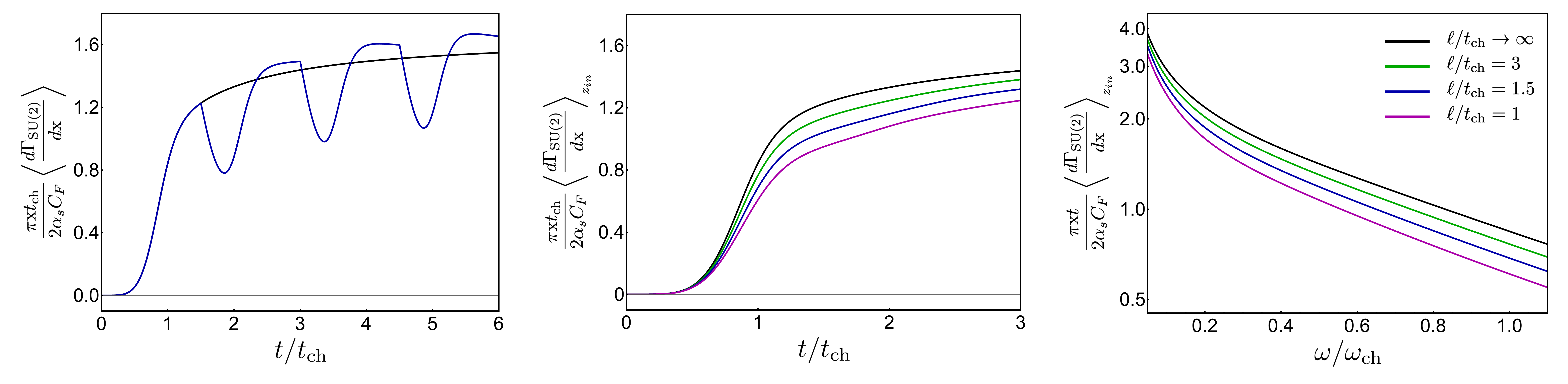}
    \caption{An illustration of the SU(2) averaged rate. The left panel gives $\frac{\pi x}{2\a_s C_f}t_{\rm ch}\left\langle\frac{d\Gamma_{\rm SU(2)}}{dx}\right\rangle$ with the leading parton starting from the edge of the first tube $z_{in}=0$. The middle and right panels depict $\frac{\pi x}{2\a_s C_f}t_{\rm ch}\left\langle\frac{d\Gamma_{\rm SU(2)}}{dx}\right\rangle_{z_{in}}$ and $\frac{\pi x}{2\a_s C_f}t\left\langle\frac{d\Gamma_{\rm SU(2)}}{dx}\right\rangle_{z_{in}}$ averaged over the starting position within the first tube. The colors correspond to the rate evaluated for different tube sizes, including the long tube limit (black), which appears in all three panels and is given by Eq.~\eqref{SU2singletube}.
    }
    \label{fig:SU2rate}
\end{figure}

Following the logic of the previous subsection, we present the numerical results for $\left\langle\frac{d\Gamma_{\text{SU(2)}}}{dx}\right\rangle$ and $\left\langle\frac{d\Gamma_{\text{SU(2)}}}{dx}\right\rangle_{z_{in}}$ in Fig.~\ref{fig:SU2rate}. As previously, the rate follows the constant field case within the first tube, experiencing some destructive interference at the edges and growing within the tubes. However, in contrast with the case of U(1), the SU(2) rate may grow somewhat faster and seems to approach the constant field curve from the above, see the left panel of Fig.~\ref{fig:SU2rate}. This behavior is featured by longer tubes, while for the shorter tubes, the rate tends to stay under the constant field curve (at least within the range of times accessible in our numerics). Varying the starting position $z_{in}$ within the first tube at a fixed tube size and averaging the rate, we find that the suppression still dominates, see the middle and right  panels of Fig.~\ref{fig:SU2rate}, which show the dependence of the rate on \(t/t_{\mathrm{ch}}\) and \(\omega/\omega_{\mathrm{ch}}\) correspondingly. Thus, we find the physical picture to be sufficiently similar to the U(1) case: as the size of the flux tubes \(l/t_{\rm ch}\) decreases, there is more destructive interference, the in-medium formation time is longer, and the rate goes down. 

\section{An alternative matter model}\label{sec:delta_model}

In this work, we rely on a simple model for the initial stages of heavy-ion collisions, which has several limitations. One of such limitations is that the averaging in Eq.~\eqref{averaging1} is centered at $E_i=0$, while the strong glasma fields are expected to be predominantly non-zero in each event. To gain further insight into the physical picture, we explore an alternative averaging procedure in this section, addressing this specific aspect of the glasma field dynamics.

To do so, we enforce the chromoelectric field of any given tube to fluctuate in its direction along the beam axis and in its color, but keeping its absolute value fixed. Thus, the alternative averaging procedure can be summarized for the SU($N_c$) case with
\begin{align}
&\langle f(E^a_{1x},E^a_{2x},E^a_{3x},\cdots) \rangle = \int_{E_1} \frac{E_0}{2}\,\Gamma\left(\frac{N_c^2-1}{2}\right)\,\d(E_{1x}-E_0)\notag\\
&\hspace{3cm}\int_{E_2} \frac{E_0}{2}\,\Gamma\left(\frac{N_c^2-1}{2}\right)\,\d(E_{2x}-E_0) \cdots f(E^a_{1x},E^a_{2x},E^a_{3x},\cdots)\,,
\label{SU2averageo}
\end{align}
where the integration runs over SU($N_c$) phase space, and we have restored the $x$ subscript. In turn, in the U(1) case, this procedure simplifies to
\begin{align}
&\langle f(E_{1x},E_{2x},E_{3x},\cdots)\rangle\notag\\
&\hspace{2cm}=\left[\prod_n\int_{E_n} \frac{dE_{nx}}{2}\big(\d(E_{nx}-E_0)+\d(E_{nx}+E_0)\big)\right]f(E_{1x},E_{2x},E_{3x},\cdots)\,,
\label{U1averageo}
\end{align}
where $E_{nx}$ is just the $x$-component of the field in the $n$th tube.

Focusing first on the jet quenching parameter, we average Eq.~\eqref{qhatWEE} using the new procedure, and readily find that
\begin{align}
\langle\hat{q}(z)\rangle_{\text{SU(2)}}=\langle\hat{q}(z)\rangle_{\text{SU(3)}}=\frac{E_0^2}{4\pi^2} (z-z_n)\,,
\end{align}
since the angular averaging over the color space of the field in the last flux tube still gives zero, unless the two fields in Eq.~\eqref{qhatWEE} are from the same flux tube. Similarly, when averaging the jet quenching parameter over the starting position $z_{in}$ in the first tube, we find
\begin{align}
\langle\hat{q}(z)\rangle_{\text{SU(2)},z_{in}}=\langle\hat{q}(z)\rangle_{\text{SU(3)},z_{in}}=\frac{E_0^2}{8\pi^2}\ell\,.
\end{align}

Turning to the U(1) medium-induced radiation, we start with the single-flux-tube emission rate. After averaging, the rate reads
\begin{align}
   &\left\langle\frac{d\Gamma_{\text{U(1)}}}{d\text{x}}\right\rangle=\frac{2\alpha_s C_F}{\text{x}\pi} \re \int_0^t ds\,  \frac{1}{s^2} \left(1 - \left(1- i\frac{E_0^2s^3}{8\omega} \right) e^{-i \frac{E_0^2 s^3}{24 \omega}}  \right)\,,
   \label{U1singletubeo}
\end{align}
Thus, one may see that $\left\langle\frac{\Gamma_{\text{U(1)}}}{d\text{x}}\right\rangle$ coincides with the constant field rate Eq.~\eqref{K1tubeNA} before averaging over multiple events, up to an identification $E=E_0$. The multiple oscillations of the rate do not average out now, and, as we will see further, they reappear for finite tube sizes.

\begin{figure}
    \centering
    \includegraphics[width=1\textwidth]{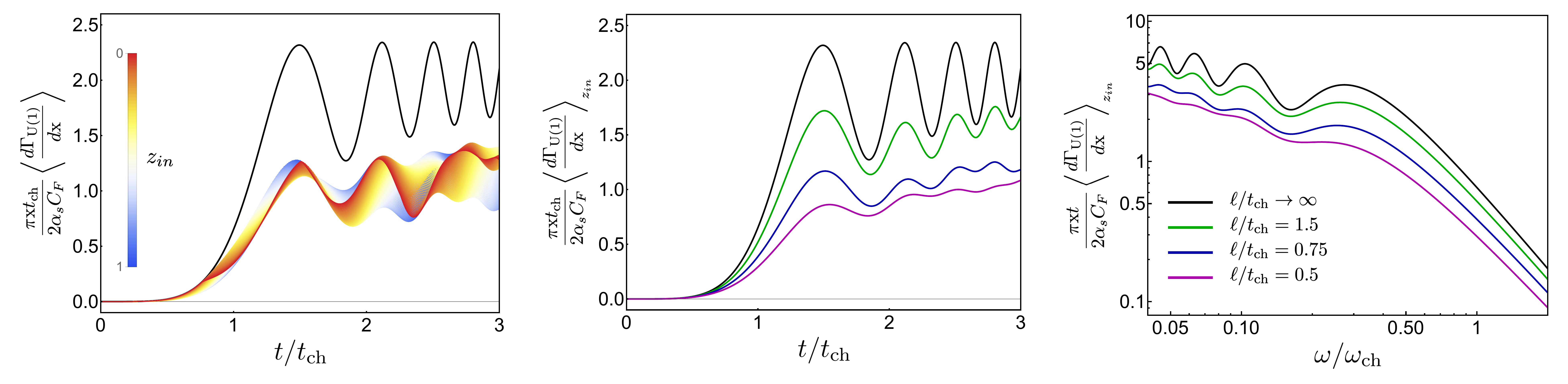}
    \caption{An illustration of the U(1) rate averaged under the alternative procedure, see Eq.~\eqref{U1averageo}. The left panel displays a family of curves for $\frac{\pi x}{2\a_s C_f}t_{\rm ch}\left\langle\frac{d\Gamma_{\rm U(1)}}{dx}\right\rangle$ at $\ell/t_{\rm ch}=0.75$ with the leading parton starting its path from different positions $z_{in}$ within the first tube. These rate curves are distinguished by the color, as indicated in the legend of the left panel. The middle and right panels depict $\frac{\pi x}{2\a_s C_f}t_{\rm ch}\left\langle\frac{d\Gamma_{\rm U(1)}}{dx}\right\rangle_{z_{in}}$ and $\frac{\pi x}{2\a_s C_f}t\left\langle\frac{d\Gamma_{\rm U(1)}}{dx}\right\rangle_{z_{in}}$ averaged over the starting position within the first tube. The colors listed in the legend of the right panel correspond to the rate evaluated for different tube sizes, including the long tube limit (black), which appears in all three panels and is given by Eq.~\eqref{U1singletubeo}. 
    }
    \label{fig:U1rateother}
\end{figure}

As before, the rate exhibits a more complex structure when the leading parton propagates through more than a single tube. Relying on the same overall parametric dependence of the averaged rate for both averaging procedures, we evaluate the multiple-tube case numerically and present our results for $\left\langle\frac{d\Gamma_{\text{U(1)}}}{dx}\right\rangle$ and $\left\langle\frac{d\Gamma_{\text{U(1)}}}{dx}\right\rangle_{z_{in}}$ in Fig.~\ref{fig:U1rateother}.
In the left panel, we illustrate the long flux-tube limit of the U(1) rate averaged under Eq.~\eqref{U1averageo}, identical to the constant field rate Eq.~\eqref{K1tubeNA} in this case. Here, we also depict a family of rates at $\ell/t_{\rm ch}=0.75$ plotted for different initial positions $z_{in}$ of the leading parton within the first tube. The corresponding curves are distinguished by a color scheme and form a colored band. The shape of this band indicates that even after averaging over the initial positions, some oscillatory patterns of the constant field rate persist. In turn, in the middle and right panels of Fig.~\ref{fig:U1rateother}, we present $t_{\rm ch}\left\langle\frac{d\Gamma_{\text{U(1)}}}{dx}\right\rangle_{z_{in}}$ and  $t\left\langle\frac{d\Gamma_{\text{U(1)}}}{dx}\right\rangle_{z_{in}}$ additionally averaged over the starting position $z_{in}$ in the first tube as functions of $t/t_{\rm ch}$ and $\o/\o_{\rm ch}$ respectively. We see that the rates averaged under the second procedure of Eq.~\eqref{U1averageo} are again ordered with $\ell/t_{\rm ch}$, falling for shorter tube sizes. 

Averaging the SU(2) rate with Eq.~\eqref{SU2averageo}, we can still rely on the Wilson line decomposition in Eq.~\eqref{Eq:Wilson_product}. Moreover, the angular averaging in the color space ensures the validity of Eq.~\eqref{WSU2averaged}. Thus, the averaged single-flux-tube emission rate in the SU(2) case reads
\begin{align}
   &\left\langle\frac{d\Gamma_{\text{SU(2)}}}{d\text{x}}\right\rangle_{\text{SU(2)}} =  \frac{2\alpha_s C_F}{3\text{x}\o^2} \re \int_0^t ds\, \left\langle\boldsymbol{\na}_\x\cdot \boldsymbol{\na}_\y \Big(\cK_+(\x,t;\y,s)+\cK_-(\x,t;\y,s)-2\cK_0(\x,t;\y,s)\Big)_{\x=\y=0}\right\rangle\notag\\
   &\hspace{1.5cm}=\frac{4\alpha_s C_F}{3\text{x}\pi} \re \int_0^t ds\,  \frac{1}{s^2} \left(1 - \left(1- i\frac{E_0^2s^3}{8\omega} \right) e^{-i \frac{E_0^2 s^3}{24 \omega}}  \right)\,,
   \label{SU2singletubeo}
\end{align}
differing from the U(1) case only by a numerical factor. 

\begin{figure}
    \centering
    \includegraphics[width=\textwidth]{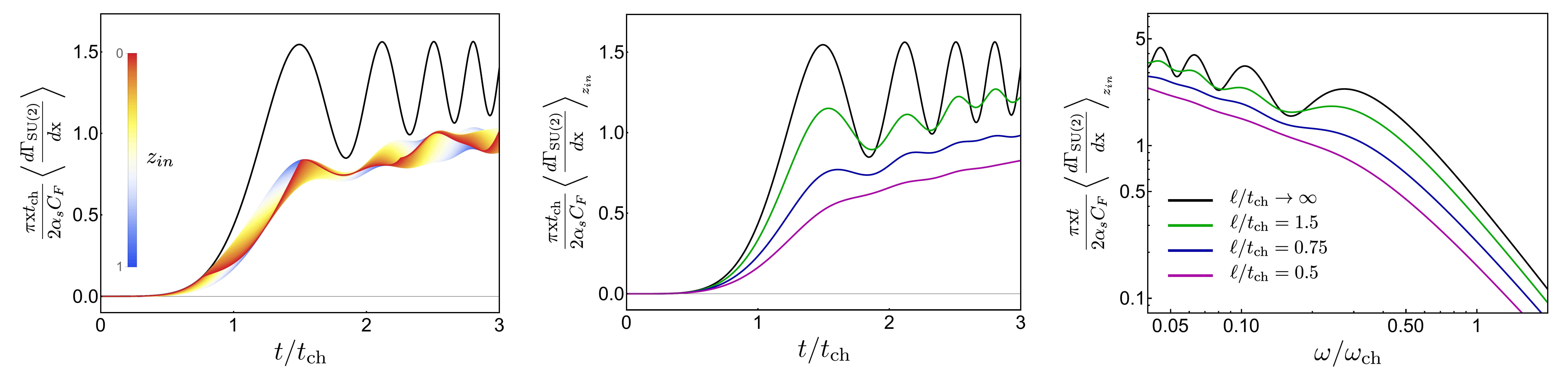}
    \caption{An illustration of the SU(2) rate averaged under the alternative procedure, see Eq.~\eqref{SU2averageo}. The left panel displays a family of curves for $\frac{\pi x}{2\a_s C_f}t_{\rm ch}\left\langle\frac{d\Gamma_{\rm SU(2)}}{dx}\right\rangle$ at $\ell/t_{\rm ch}=0.75$ with the leading parton starting its path from different positions $z_{in}$ within the first tube. These rate curves distinguished by the color, as indicated in the legend of the left panel. The middle and right panels depict $\frac{\pi x}{2\a_s C_f}t_{\rm ch}\left\langle\frac{d\Gamma_{\rm SU(2)}}{dx}\right\rangle_{z_{in}}$ and $\frac{\pi x}{2\a_s C_f}L\left\langle\frac{d\Gamma_{\rm SU(2)}}{dx}\right\rangle_{z_{in}}$ averaged over the starting positions within the first tube. The colors listed in the legend of the right panel correspond to the rate evaluated for different tube sizes, including the long tube limit (black), which appears in all three panels and is given by Eq.~\eqref{SU2singletubeo}.
    }
    \label{fig:SU2rateother}
\end{figure}

Similarly to the U(1) case, we focus on the numerical estimates of the SU(2) rate at finite tube length, presenting the results in Fig.~\ref{fig:SU2rateother}. The left panel again exhibits the long tube limit of the SU(2) rate averaged under Eq.~\eqref{SU2averageo} along with a family of the SU(2) rates at $\ell/t_{\rm ch}=0.75$ with different starting positions $z_{in}$ of the leading parton within the first tube (distinguished by a color scheme). In turn, the middle and right panels depict $t_{\rm ch}\left\langle\frac{d\Gamma_{\text{SU(2)}}}{dx}\right\rangle_{z_{in}}$ and $t\left\langle\frac{d\Gamma_{\text{SU(2)}}}{dx}\right\rangle_{z_{in}}$ averaged over the initial position within the first tube, and one may see the same ordering of the rates with $\ell$.

Thus, we conclude that, while certain nuances of the rate may vary depending on the specific matter model and gauge group, features such as the anticipated asymptotic behavior and qualitative dependence of the rate on $\ell/t_{\rm ch}$ stay untouched, indicating sufficient robustness of the physical picture obtained in this work.

\section{Conclusions}\label{sec:conclusion}

In this work, we have developed a formalism to describe jet quenching during the early times of heavy-ion collisions, when the medium is dominated by strong color fields. Throughout our considerations, we have relied on a simplified matter model that captures the main features of this glasma phase. In particular, we have assumed that the matter is made up of color domains with chromoelectric fields confined in flux tubes aligned with the beam axis. The field is homogeneous and constant within any given tube, and independently varies in different domains on an event-by-event basis. Thus, we relax several assumptions that are ubiquitous in most of the jet quenching considerations. First, we go beyond the approximation of local field correlations widely used in the jet quenching claculations in the quark-gluon plasma context,\footnote{The recent developments of the jet quenching theory in anisotropic media may provide another example of going beyond the limit of local correlations in the transverse directions, see e.g. \cite{Barata:2022utc}.} and assume a finite correlation length. This is also in contrast with the constant (slowly-varying) field limit employed in considerations of the synchrotron-like radiation of energetic partons. Second, we consider the event-by-event fluctuations of the glasma background and its interplay with the jet physics, providing a starting point for jet tomography of the early stages of heavy-ion collisions. 

Using our model, we start by calculating the transverse momentum broadening of jet partons for both SU(2) and SU(3) backgrounds, investigating how the results depend on the size of the color domains $\ell$. The jet quenching parameter for a jet parton traversing the given flux tube increases with time. Conversely, for a jet parton traversing many small flux tubes, the stochastic kicks it receives average out, leading to a constant rate of broadening.

We further focus on the medium-induced radiation in the glasma background, restricting our consideration to the emission rate in the soft gluon limit. Most of our results are obtained for a SU(2) background, where the model can be treated semi-analytically, and compared with the toy consideration for the U(1) case. However, it should be emphasized that the presented formalism can be straightforwardly extended to the physical SU(3) case. We have also explored two different averaging procedures -- a Gaussian averaging centered at zero field and a delta-functional averaging with fluctuating direction -- to probe the sensitivity of our results to the features of the medium model.

The behavior of the rate can be understood by comparing the relevant timescales of the problem. When a single flux tube is longer than the formation time for gluon emission, the resulting rate reproduces the well-known form for synchrotron radiation in a constant field. Considering the characteristic energy scales for the saturation and jet quenching processes in heavy-ion collisions, one may expect this regime to be relevant only for very soft emissions. At higher energies, the partons traverse multiple flux tubes during the emission process. This leads to an intricate interplay between different flux tubes, due to both momentum kicks in different directions and color decoherence between flux tubes. The net result is a reduction in the rate when the flux tubes are smaller, which is especially pronounced for higher gluon energies or at later times. This result holds both for U(1) and SU(2) rates, and for both averaging procedures, as illustrated in Fig.~\ref{fig:U1rate}, Fig.~\ref{fig:SU2rate}, Fig.~\ref{fig:U1rateother}, and Fig.~\ref{fig:SU2rateother}.

Finally, we study the details of the in-medium formation time behavior for jet quenching in our medium model. To do this, we consider the rate as a function of energy and tube size at fixed $\hat{q}$ and $t$. The similarity of the SU(2) and U(1) rates (under two averaging procedures) suggests that such features are sufficiently universal, and we focus on the U(1) case with Gaussian averaging for simplicity, see Fig.~\ref{fig:fixedqhat}, leaving more general investigation for future. The U(1) rate exhibits three distinct scaling regimes: \textbf{(1)} At lower energies the partons traverse only a few tubes, and the rate scales as $\o^{-1/3}$ resembling the behavior in the case of a single long tube; \textbf{(2)} For the transition region of intermediate energies, the partons traverse multiple tubes during the emission process, but the formation time is smaller than $t$. The rate scales as $\o^{-1/2}$, recovering the behavior of the BDMPS-Z rate under the harmonic approximation. This is well-expected, since in this limit the finite correlation length $\ell$ is small, and averages are near local; \textbf{(3)} For larger energies the formation time is larger than $t$, and the rate scales as $\o^{-2}$, further recovering the large energy behavior of the BDMPS-Z rate under the harmonic approximation. The latter region goes beyond the applicability of the harmonic approximation in the BDMPS-Z case, and it is taken over by the small number of hard interactions of the opacity expansion regime, see e.g. \cite{Sievert:2018imd} for a review. It would be interesting to further study whether similar larger transverse momentum exchanges may take place in the glasma phase and investigate their relevance. We leave such a study for future work. 

The presented results can be extended in several ways to achieve a more realistic description of jets during the early stages of heavy ion collisions. A natural and promising future step would be to combine our approach with realistic numerical simulations of the glasma. This would allow us to compute the medium-induced soft gluon rate, similarly to the previous studies on single particle momentum broadening \cite{Ipp:2020mjc, Ipp:2020nfu, Avramescu:2023qvv}. In particular, such an exercise would help determine if the early time evolution of jets is indeed described by a multiple scattering picture, in the style of the BDMPS-Z formalism. Beyond this application, in order to compute realistic jet observables, our model needs to be complemented with the virtuality cascade of the jet evolution and account for the subsequent stages of matter evolution. 

\begin{acknowledgments}
We thank F. Gelis, E. Iancu, C. Salgado, and B. Schenke for helpful discussions related to this work. JB is supported by the United States Department of Energy under Grant Contract DESC001270. SH is grateful to the EIC Theory Institute and to Brookhaven National Laboratory for the hospitality and support during the early stages of this work. XML is supported under scholarship No. PRE2021-097748, funded by MCIN/AEI/10.13039/501100011033 and FSE+. XML is also supported by Xunta de Galicia (Centro singular de investigación de Galicia accreditation 2019-2022), by European Union ERDF; and by Grant CEX2023- 001318-M funded by MICIU/AEI/10.13039/501100011033 and by ERDF/EU. AVS is supported by Funda\symbol{"00E7}\symbol{"00E3}o para a Ci\symbol{"00EA}ncia e a Tecnologia (FCT) under contract 2022.06565.CEECIND. AVS and XML also acknowledge the support from European Research Council project ERC-2018-ADG-835105 YoctoLHC. This work has been also supported by STRONG-2020 "The strong interaction at the frontier of knowledge: fundamental research and applications” which received funding from the European Union’s Horizon 2020 research and innovation program under grant agreement No 824093. 

\end{acknowledgments}

\bibliographystyle{bibstyle}
\bibliography{references}

\end{document}